# An Analysis of Nanoparticle Settling Times in Liquids

D D Liyanage[1], Rajika J K A Thamali[1], A A K Kumbalatara[1], J A Weliwita[2], S Witharana[1,3] *

[1]Department of Mechanical and Manufacturing Engineering, University of Ruhuna, Galle, Sri Lanka

[2]Department of Mathematics, University of Peradeniya, Kandy, Sri Lanka

[3]Sri Lanka Institute of Nanotechnology, Colombo, Sri Lanka

*Corresponding author: switharana@ieee.org

**ABSTRACT**:

Aggregation and settling are crucial phenomena that involve particulate systems. For particle sizes of millimetre and above, there are reasonable accurate predicting tools. However for smaller particle sizes, there appears to be a void in knowledge. This paper presents an analytical model to predict the settling rates of nano-to-micro size particulate systems. The model was developed as a combination of modified classical equations and graphical methods. A calculation sequence also is presented. By validation with available experimental data for settling nano-to-micro systems, it was found that the two schemes show order of magnitude agreement. A significant feature of this model is its ability to accommodate non-spherical particles and different fractal dimensions.

**INTRODUCTION**

Nanoparticles, for their remarkable magnetic, optical, thermal and transport properties, have drawn interest from scientific communities across a broad spectrum. Among these, the thermal conductivity is an important property that has been extensively investigated and debated over the past twenty years. Addition of traces of nanoparticles, often less than 1 vol% to a common heat transfer liquid, has enabled increasing the thermal conductivity of the same liquid by up to 40% or more [1-3]. However there are a few contradictory observations too. In the famous INPBE experiment [4], it was observed that the measured thermal conductivity enhancements did not surpass the range predicted by the classical effective medium theories. Nanoparticle-liquid heat transfer blends are nowadays popularly known as *nanofluids*.

Particles suspended in liquids have the tendency to form aggregates that would finally lead to separation and gravity settling. Aggregation is also contemplated as a major mechanism responsible for the enhanced thermal conductivity of nanofluids [5]. A settling nanofluid however would be of little practical use. In addition to the underperformance in thermal conductivity, separated large aggregates may clog filters and block the flow in narrow pipelines in machines. However, for mineral extraction and effluent treatment industries, separation and settling are basic prerequisites of operation. To welcome or to avoid it, one may need to understand the settling processes and settling rates. Despite its critical importance, only a few experimental studies have been dedicated to explore the aggregation and settling dynamics of nanoparticles [6]. Complexity of aggregating nanoparticulate systems [6-8] and difficulties in taking accurate measurements seem to be major challenges against the progress of experimental work.

This paper presents analytical method to calculate the nanoparticle aggregation and settling times in liquids. To start with, one should know the particle concentration in the nanofluid. In a rare case of unknown particle concentration, one could measure the viscosity of the nanofluid and compute it, as suggested by Chen et al [9]. Second step is to determine the aggregation time. A correlation for aggregation time is derived by considering all governing parameters. As the third step, a procedure to determine settling time is proposed. This procedure is a combination of equations, graphs and estimations. The total time for settling should be the sum of the aggregating time and the settling time determined this way.

Lastly the proposed model is validated by comparing the calculated values with the limited experimental data available in literature.

**MATERIALS AND METHODS**

**I. Aggregation Model**

Consider a system where the nanoparticles are well dispersed in a liquid. Aggregates will start to form driven by a number of parameters such as nanoparticles size, concentration, solution temperature, stability ratio, and fractal dimensions [20-22]. At any given time $t$, these aggregates can be characterized by radius of gyration ($R_a$);

$$\frac{R_a}{r_p} = (1 + t/t_p)^{1/d_f} \quad \text{..........................} (1)$$

where $d_f$ is the fractal dimension of the aggregates, practically found to be between 2.5-1.75[23].

The aggregation time, $t_p$ is defined as[8];

$$t_p = (\pi \mu r_p^3 W)/(k_B T \emptyset_p) \quad \text{........................} (2)$$

Here $K_B$, $T$, $\emptyset_p$ and $\mu$ are respectively the Boltzmann constant, temperature, particle volume fraction, and viscosity of the liquid.

The stability ratio $W$ in the equation (2) is defined by,

$W = 2r_p \int_0^\infty B(h) exp\left\{\frac{V_R+V_A}{k_BT}\right\}/(h+2r)^2 dh$ where $B(h)$ is the factor that counts the hydrodynamic interaction and according to Honing et al[24] for inter particle distance $h$,

$$B(h) = \frac{6(h/r_p)^2 + 13\left(\frac{h}{a}\right) + 2}{6(h/r_p)^2 + 4\left(\frac{h}{a}\right)}.$$

Moreover, $V_R$ is the repulsive potential energy and is given by

$V_R = 2\pi\epsilon_r\epsilon_0 r_p\psi^2 exp(-\Lambda h)$ for dielectric constant of free space, $\epsilon_0$ and $\zeta$ potential, $\psi$. Note that this is valid when $\Lambda r_p < 5$.

Debye parameter, $\Lambda = 5.023 \times 10^{11} I^{0.5}/(\epsilon_r T)^{0.5}$ where $\epsilon_r$ is the relative dielectric constant of the liquid and $I$ is the concentration of ions in water that can be related to the pH in the absence of salts, by $I = 10^{-pH}$ for pH $\leq$ 7 and $I = 10^{-(14-pH)}$ for pH > 7.

Further the attractive potential energy is defined as

$V_A = -A/6\left[\frac{2r_p^2}{h(h+4r_p)} + \frac{2r_p^2}{(h+2r_p)^2} + \ln(h(h+4r_p)/(h+2r_p)^2)\right]$, where $A$ is the Hamaker constant

Using above set of equations, one could compute the time taken ($t_p$) to form an aggregate of a known radius ($R_a$).

## II. Settling Time

Terminal settling velocity of a particle in motion in a fluid body depends is governed by multiple factors such as fluid density, particle density, size, shape and concentration, degree of turbulence and temperature etc. [25]. The Stokes law of settling was originally defined for small, $mm$ or $\mu m$ size spherical particles with low Reynolds numbers ($Re \leq 0.3$). The drag force of a creeping flow over a rigid sphere consists of two components, viz., pressure drag ($Fp = \pi\mu dU$) and the shear stress drag ($F = 2\pi\mu dU$)[26-28]. Thus the total drag becomes $FD = 3\pi\mu dU$. Using the Stokes equation for a spherical particle in ideal conditions; infinite fluid volume and in lamina flow and zero acceleration,

$$\frac{4\pi R^3 (\rho_p - \rho_f) g}{3} - 6\pi\mu UR = 0 \ldots \ldots \ldots \ldots (X)$$

where $U$ is terminal velocity and $R$ is the equivalent radius of the aggregate. Stokes drag ($6\pi\mu UR$) could be re-expressed as, $F_{sd} = C_D \frac{\rho_f U^2}{2} A$, where $C_D = \frac{F'}{\frac{1\rho_f U^2}{2}}$ for force $F'$ per unit projected area, and $A$ is the projected area of the particle to on coming flow and for a sphere, $A = \frac{4\pi R^3}{3}$. $C_D$ is found from figure 3. Moreover, a particle in a creeping flow where Reynolds number is very small tend to face the least projected area to the flow[27].

Thus the equation *(X)* becomes, $\frac{4\pi R^3 (\rho_p - \rho_f) g}{3} - C_D \frac{\rho_f U^2}{2} A = 0$ and hence,

$$U = \sqrt{\frac{8\pi R_a^3 (\rho_p - \rho_f) g}{3 C_D \rho_f A}} \ldots\ldots\ldots\ldots\ldots\ldots\ldots\ldots\ldots\ldots (3)$$

However, a nanoparticle will hardly qualify for the Stokes conditions because of its larger surface area to volume ratio. In such circumstances, the surface forces dominate over gravitational forces. Moreover for a nanoparticle dispersed in a fluid, the intermolecular forces

(Van der Waal's, Iron-Iron interactions, Iron-Dipole interactions, Dipole-Dipole interactions, Induced Dipoles, Dispersion forces and overlap repulsion) [29, 30] along with the thermal vibrations (Brownian motions) and diffusivity will take over the Newtonian forces.

Experimental results show the nanoscale (*1nm-20nm*) particles form microscale (*0.1-15 μm*) aggregates [23]. The shape of the aggregates depends upon fractal dimension ($d_f$)that varies between 1.5-2.5 in most cases. When $d_f$ approaches 3, the shape of aggregates get closer to a spherical shape. Also when a colony of nanoparticles form one microsize aggregate, the intermolecular forces disappear permitting the application of Newtonian forces on the aggregate [31]. For instance, mean free path ($L = \sqrt{\frac{2kT}{3\pi x \mu}} \, t$) due to Brownian motion shortens by 86.44% when nanoparticles of $23nm$ are clustered together and formed a $2.5\mu m$ aggregate [32]. Following are the assumptions and conditions for application Stokes law for the aggregate model.

*(a). Reynolds number*

Aggregates have very low settling velocities, thus they give very small Reynolds numbers. For example, *Re* can be expected in the region of 0.001~0.0001 [7].

$$\mathrm{Re} = \frac{D_a \, U \, \rho_f}{\mu} \quad \ldots\ldots\ldots\ldots\ldots\ldots\ldots\ldots\ldots\ldots\ldots\ldots\ldots\ldots\ldots\ldots\ldots\ldots\ldots \quad (4)$$

Derivation of *Re* of settling aggregates is not straightforward since only their diameters are known but not the settling velocities. An iterative method is therefore proposed[28].

$$C_D \, Re^2 = \frac{4g(\rho_p - \rho_f)\rho_f d^3}{3\mu^2} \quad \ldots\ldots\ldots\ldots\ldots\ldots\ldots\ldots\ldots\ldots\ldots \quad (5)$$

This enables determination of *Re* using the aggregate diameter as shown in Figure 1.

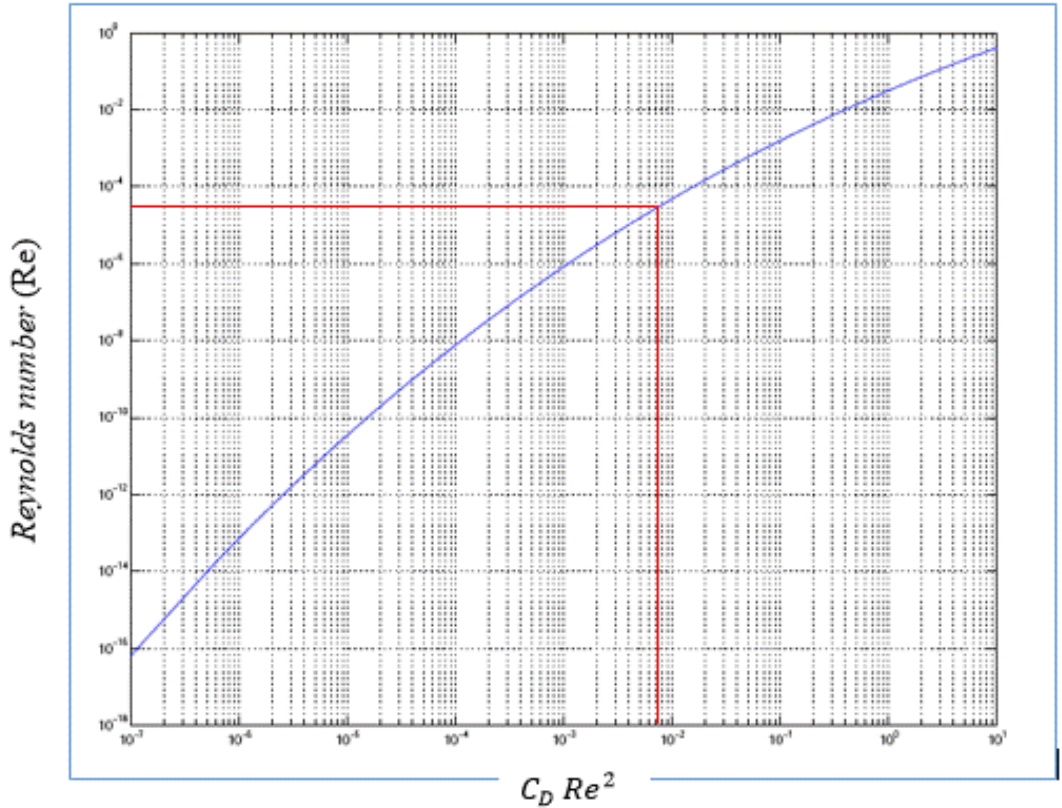

**Figure 1:** $Re$ versus $C_D Re^2$ for spherical particles.

*(b). Sphericity (ψ)*

Sphericity (ψ) indicates how spherical the aggregate is.

$$Sphericity\ (\psi) = \frac{surfce\ area\ of\ a\ sphere}{surface\ area\ of\ the\ aggregate\ which\ has\ the\ same\ volume\ of\ the\ sphere} \quad \ldots\ldots (6)$$

Rhodes [27] developed graphs to correlate the $Re$ and $C_D$ for different values of sphericity (ψ). These graphs are shown on figure 2. However the aggregates of smaller sizes have very low Reynolds numbers in the order of $10^{-4} \sim 10^{-6}$. In order to obtain $C_D$ for these low Reynolds numbers, graphs on figure 2 were extended to the left hand side and presented separately on figure 3. Equations corresponding to the set of graphs are listed on table 1.

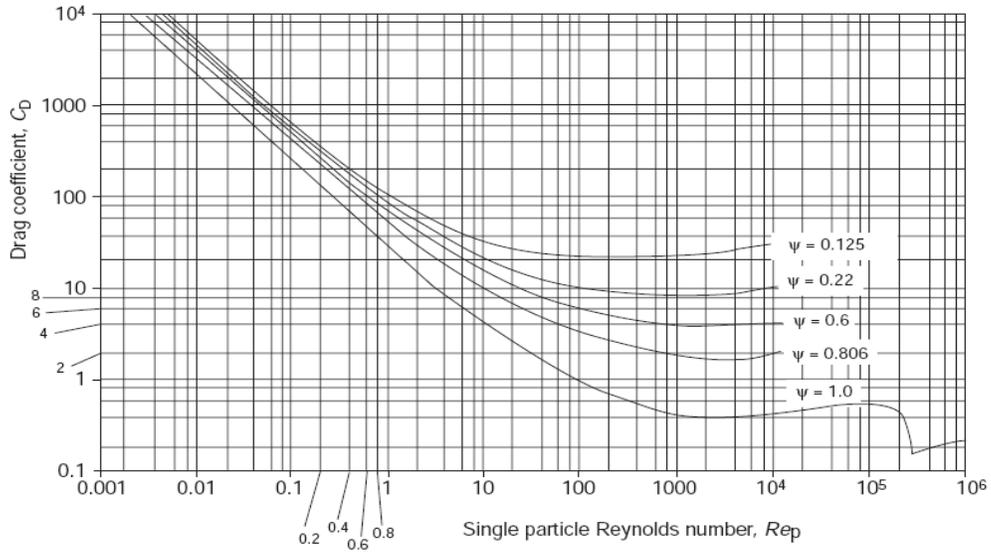

**Figure 2:** Drag coefficient ($C_D$) and Reynolds Number ($Re$) for different sphericity (ψ)[27]

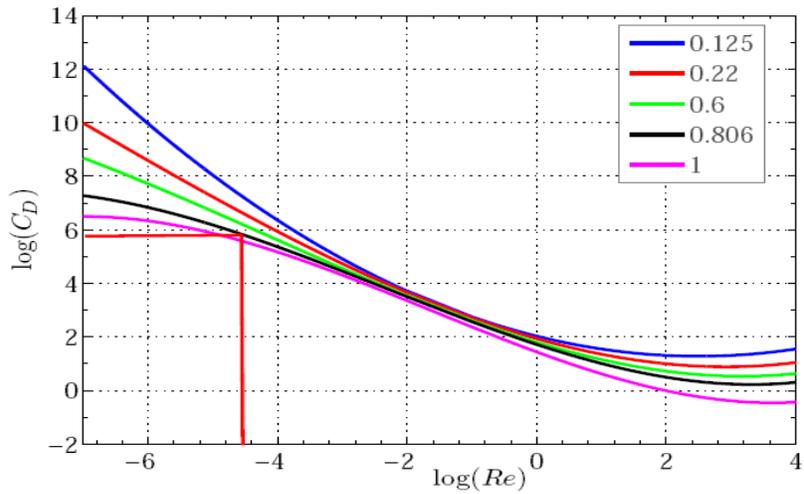

**Figure 3:** Drag coefficient ($C_D$) for low $Re$ for different $\psi$.

**Table 1:** Equations of logarithm of Drag coefficients ($C_D$) with respect to logarithm of $Re$ for different values of sphericity ($\psi$)

| $\psi$ | Relationship of Drag coefficient ($C_D$) and low Reynolds numbers ($Re$) |
|---|---|
| 0.125 | $\log(C_D) = 0.1202 \log(Re)^2 - 0.6006 \log(Re) + 2.032$ |
| 0.22 | $\log(C_D) = 0.0043161 \log(Re)^3 + 0.9725 \log(Re)^2 - 0.68 \log(Re) + 1.937$ |
| 0.6 | $\log(C_D) = 0.0067 \log(Re)^3 + 0.083 \log(Re)^2 - 0.73 \log(Re) + 1.8$ |
| 0.806 | $\log(C_D) = 0.0099 \log(Re)^3 + 0.0697 \log(Re)^2 - 0.7906 \log(Re) + 1.7225$ |
| 1 | $\log(C_D) = = 0.0116 \log(Re)^3 + 0.05793 \log(Re)^2 - 0.8866 \log(Re) + 1.443$ |

To use figure 3, one needs to know the sphericity ($\psi$) of aggregates. This information is given on table 2 for common general shapes [22, 33, 34].

**Table 2:** Geometric details of different aggregate shapes. $R_1$, $R_2$, $R_3$ are respectively aggregates of radii 2.5μm, 4μm and 5μm.

| Shape | | Volume (μm3) | Surface area (μm2) | Projected cross-section area (μm2) | $d_f$ | Sphericity ($\psi$) |
|---|---|---|---|---|---|---|
| 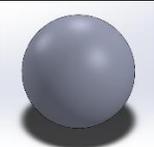 | $R_1$ | 65.45 | 78.54 | 19.63 | 3 | 1 |
| | $R_2$ | 268.08 | 201.06 | 50.26 | 3 | 1 |
| | $R_3$ | 523.60 | 314.16 | 78.53 | 3 | 1 |
| 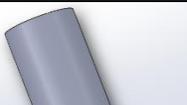 | $R_1$ | 65.44 | 91.62 | 6.61 | 1.8-2.0 | 0.857 |
| | $R_2$ | 268.10 | 293.24 | 12.57 | 1.8-2.0 | 0.686 |
| | $R_3$ | 523.08 | 388.60 | 33.18 | 1.8-2.0 | 0.808 |
| 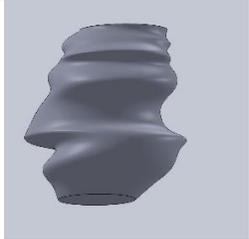 | $R_1$ | 65.43 | 110.51 | 16.14 | 2.0-2.5 | 0.710 |
| | $R_2$ | 268.10 | 294.76 | 72.27 | 2.0-2.5 | 0.682 |
| | $R_3$ | 523.62 | 419.00 | 92.56 | 2.0-2.5 | 0.75 |

*(c). Density of the aggregate*

Smoluchowski model states that nanoparticles clustered together forms a complete sphere with voids inside [8]. Moreover, when fractural dimension decreases, the aggregate geometry gets closer to a two dimensional flat object. In this case $d_f$ approaches 1.8 and appears like figure 3(a).

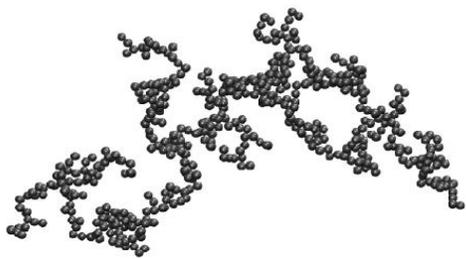

Figure 3 (a): model of an aggregate [23]

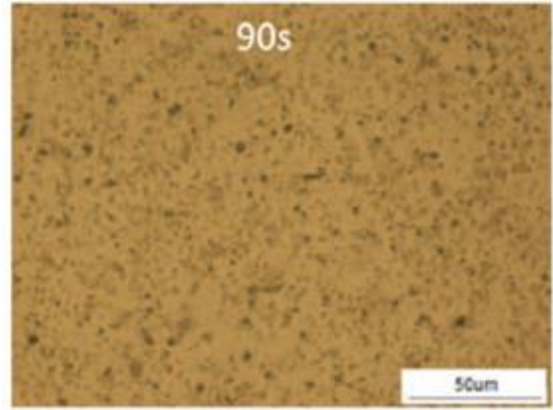

Figure 3 (b): Optical microscopy images of settling $Al_2O_3$ aggregates

Consider density of a settling aggregate. This flat object is surrounded by a thin layer of liquid molecules. Given the comparatively large density of the solid, the density of the settling aggregate can safely be assumed equal to the density of the solid.

*(d). Density of homogeneous solution*

In preparation of nanofluids, the suspension is stirred for particles to evenly distribute in the container. Along the same lines, here it is assumed that the aggregates too are evenly distributed throughout the liquid. Thus this becomes homogenous flow of aggregates. The density of the solid-liquid mixture is given by,

$$\rho_m = \emptyset_a \rho_a + (1 - \emptyset_a)\rho_f \quad \text{...................................... (7)}$$

Were $\emptyset_a$ is the aggregate volume fraction given by,

$$\emptyset_a = \emptyset_p \left(\frac{R_a}{r}\right)^{3-Df} \quad \text{................................................ (8)}$$

*(e). Viscosity of the liquid*

For volume concentrations less than 5% Einstein [36] found the following relationship using the phenomenological hydrodynamic equations.

$$\mu = \mu_0(1 + 2.5\emptyset_p) \quad \text{............................................... (9)}$$

*(f). Zero slip condition & smooth surface*

When nanoparticles are dispersed in water, the water molecules make an orderly layer around the particle, a phenomena known as liquid layering [9,35]. The water layer directly touching the particle gets denser than the bulk liquid further away. Due to this particle-water bond, it is reasonable to assume a no slip region for water. Further, the surface of the aggregate is smooth and therefore the drag due to roughness of the aggregate may not come into effect [37].

$F_A = 6\pi\mu_b R_a U_a (\frac{2\mu_b + R_a\beta_{ab}}{3\mu_b + R_a\beta_{ab}})$, where $\mu_b$ is viscosity of the fluid and $\beta_{ab}$ is *coefficient of sliding friction*. When there is no tendency for slipping $\beta_{ab} \approx \infty$ and therefore the above expression becomes the Stokes law again. $F_A = 6\pi\mu_b R_a U_a$. Hence the equation *(X)* can be used.

*vii. Batch settling*

Original Stokes theory is for a sphere travelling in an infinite medium. However in the aggregation and settling systems studied in this work, they are in large number in a finite volume of liquid. Those close proximity aggregates are influenced by each other. Richardson and Zaki [26] defines batch settling velocity ($U_p$) or particle superficial velocity. When Re < 0.3, ($U_p$) = $U_T \varepsilon'^{4.65}$ where $\varepsilon'$ is liquid volume fraction, $\varepsilon' = \frac{Voids\ Volume}{Total\ Volume}$. Here $\varepsilon'$ should be less than 0.1. In nanofluids however the particle concentrations are far smaller than this. For example, the liquid volume fraction in the Witharana et al [7] settling experiment, $\varepsilon'$ was 0.723 ($\varepsilon' = 1 - \emptyset_p (\frac{R_a}{r})^{3-Df}$). Hence in the context of this work, the batch settling scenario is very weak.

**RESULTS AND DISCUSSION**

**Determination of aggregation and settling times from the proposed model**

For the validation of this model, the experimental data from Witharana et al [7] were recruited. Their system was polydisperse spherical alumina ($Al_2O_3$) nanoparticles suspended in water at near-IEP. The sizes were ranging between 10~100nm, with the average size of 46*nm*. From the optical microscopy images aggregates are seen to have radius between $1\mu m$ ~$10\mu m$. (figure 3(b)). For validation therefore the equivalent aggregate radius is taken as 2.5$\mu m$, and density of $Al_2O_3$ is taken as 3970kg/m³. Fractural dimension ($d_f$) is assumed to be

1.8, based on the geometry of the aggregate shown on the microscopy images. Their nanoparticle concentration was 0.5 wt% which converts to equivalent volume fraction ($\phi_p$) of 0.001 vol%. The height of the vials where the samples were stored during the experiment was 6cm.

*Aggregation time*

At IEP, the repulsive and hydrodynamic forces become minimum and the value of W tends to 1. Now using equations (1) and (2), $t_p$ and aggregation time *t* is calculated for the following values:

$\mu = 8.92 * 10^{-4}$ kg/m/s, $r_p = 23 nm$, $\phi_p = 0.001$, $T = 293 K$ and $k_B = 1.38 * 10^{-23}$ J/m²/K⁴/s

$t_p = 8.45 \times 10^{-3}$ s  Where $R_a = 2.5 \mu m$ and $d_f = 1.8$

This yields an aggregation time t = 0.65mins *(a)*

*Settling velocity*

- Density of the aggregate ($\rho_{agg}$) is taken from above part 3, which is 3970kg/m³.
- Density of homogenous flow ($\rho_m$) is calculated from equation (7), which is 1003 kg/m³
- Aggregation fraction ($\phi_a$) is calculated from equation (8), which is 0.277 and Void fraction of the liquid ($\varepsilon'$) is 0.723.
- Viscosity of the liquid ($\mu$) is calculated from equation (9), which is 8.92×10⁻⁴kgfsm⁻²
- Calculated value for $C_D Re^2$ for 5µm diameter of aggregate from equation (5) is 6.14×10⁻³
- From the figure 1 we can then get the *Re* approximately as $3\times10^{-5}$.
- Spherical (ψ) is taken as 0.857 from the table 1, hence Spherical (ψ) is in between 1 and 0.806

- From the graph in figure 3 Drag coefficient ($C_D$) is calculated to be approximately $3 \times 10^5$. (shown by arrows)

- From the equation (3) and taking equivalent radius as 2.5µm, terminal velocity is calculated to be $4.37 \times 10^{-5}$ m/s *(b)*.

- On the experimental study [7], the actual settling velocity was reported as $6.66 \times 10^{-5}$ m/s. This falls within the same order of magnitude as calculated in the previous step *(b)*.

*Total settling time*

Total time for settling should be equal to the sum of aggregation time and settling time. Aggregation time was calculated above *(a)* as 0.65 mins. Once aggregates were formed, assume they reached the terminal velocity in negligible time. Now the total time for settling becomes,

Total time for settling = aggregation time + settling time

$$= 0.65 min + \frac{6 \times 10^{-2}}{4.37 \times 10^{-5}}$$

$$= 0.65 + 22.88$$

$$= 23.53 min$$

Figure 3 in Witharana et al [7] provides the pictures of their settling nanofluid. Close to 30 mins after preparation, the samples were fully settled. Calculation presented above is therefore is in good agreement with the actual experiment.

**CONCLUSIONS**

Determination of the settling rates of nano and micro particulate systems were of both academic and industrial interest. To analyze these complex systems, experimentation would be the most accurate method. However for most practical applications, predictability of settling rates is of utmost importance. The equations available in literature address the sizes of sub-millimeter or above. Thus there was a gap for a predictive model that can cater to nano and micrometer sized particles. The work presents in this paper was an effort to fill this gap. To begin the procedure, one needs to know the particle concentration in the liquid. Then, first the aggregation rates were calculated using modified correlations. Settling rates were then determined from a combination of equations and graphs. To validate this model, experimental data for $Al_2O_3$–water system was recruited from literature. Their settling rates were $6.66 \times 10^{-5}$ m/s whereas the model prediction was $4.37 \times 10^{-5}$ m/s. Thus the analytical and experimental schemes were in reasonable agreement to the same order of magnitude.

The versatility in this model is that it can accommodate roundness deviations and fractal dimensions ($Df$). However, for further validation of the model and fine tuning, more experimental data are required.